# Theoretic Guide for Using Photonic Glasses as Colored Covers for Solar Energy Harvesting

Zhenpeng Li, Sinan Li, Tao Ma*

Engineering Research Centre of Solar Energy and Refrigeration of MOE, School of Mechanical Engineering, Shanghai Jiao Tong University, Shanghai 200240, People's Republic of China

*Corresponding author: Tao Ma (tao.ma@sjtu.edu.cn)

**Abstract**

The increasing demand for renewable energy is promoting technologies that integrate solar energy harvesting materials with the human living environment, such as building-integrated photovoltaics (BIPVs). This places requirements on developing colored covers with a trade-off between efficiency and aesthetics, providing a new stage for the large-scale application of structural color technologies. Here in this study, we have investigated the theoretic feasibility of employing the photonic glass, a random packing of monodisperse dielectric microspheres, as the colored cover for solar energy harvesting. Based on numerous optical simulations, we have evaluated the color and average solar transmissivity (AST) of the photonic glasses with varying parameters. Results show that using non-absorbing microspheres with relatively high refractive index, about 3 μm thick photonic glasses could enable colors with lightness over 50 while keeping AST at around 80%. Besides, we demonstrate that due to the short-range structural correlation, photonic glasses could generate purple, blue, cyan, light green, and grey colors, but cannot help with yellow and red color hues. Finally, the effects of several enhancement methods are clarified, and possible ways for expanding the color range are demonstrated. These results provide a comprehensive guide to the practical implementations of structural color using photonic glasses, particularly in the colorization of solar energy materials.

Key words: photonic glass, disordered photonics, colloidal assembly, structural color, solar energy

# Introduction

Due to the low intensity, using solar energy to power a sustainable future requires large areas of land. But the land is a scarce resource in the human living environment, particularly in cities. This stimulates great interest in technologies that could integrate solar energy harvesting into our daily life on a large scale, for example, building-integrate thermal collector and building-integrated photovoltaic (BIPV).[1, 2] For maximizing efficiency, the most typical feature of artificial solar energy harvesting devices like solar thermal (ST) collectors and solar photovoltaics (PV) is their nearly black appearance. However, a black wall or roof is unpopular with most people, therefore in this scenario, attractive aesthetics are just as important as high efficiency.[3-6] This places great demand on developing colorful cover plates for ST and PV, which should promise a decent trade-off between efficiency and aesthetics, and thus should have the following core features: 1) selective visible light (VIS) reflection; 2) negligible solar radiation absorption; 3) high near-infrared light (NIR) transmission.[7]

Unfortunately, due to their absorption of non-reflected VIS or strong NIR reflection, common colorant materials like pigments and dyes are generally not compliant.[8] In contrast to color by selective absorption, structure color is generated by the scattering, diffraction, and interference of light in micro- and nano-structures.[9-11] This makes it possible to achieve colors covering the entire visible spectrum with solar-transparent materials.[12-14] However, although periodic photonic structures with long-range order could show strong selective reflection and vivid colors, the production process is generally material-consuming and time-consuming, preventing large-scale applications. In this context, the amorphous structural color might be a better candidate, which requires only random structures with short-range correlations.[15-17] Besides, the isotropic structure makes color angle-independent, and appears pretty close to that produced by absorbing pigments and dyes.[18, 19] In recent years, the artificial construction of amorphous structural color materials has seen rapid development, especially those made by self-assembled dielectric microspheres, referred to as photonic glasses.[20-22]

Previous studies have demonstrated the feasibility of replacing traditional pigments with photonic glasses or photonic glasses with light-absorbing additives, in which the following advantages are often highlighted: long-term stability, environmental friendliness, and dynamic tunability.[23-25] While in most cases, another advantage of all-dielectric photonic glasses for colorization is not explored, i.e., negligible absorption and high NIR transmission. In this context, we envisioned that photonic glasses and solar energy harvesting materials would be a win-win combination. Recently we reported mass-producible and high-efficiency colored PVs using the photonic glass self-assembled by colloidal ZnS microspheres, preliminarily validating the idea.[26] To guide practical applications in the future, the relationship between structure, color, and solar radiation transmittance needs to be figured out.

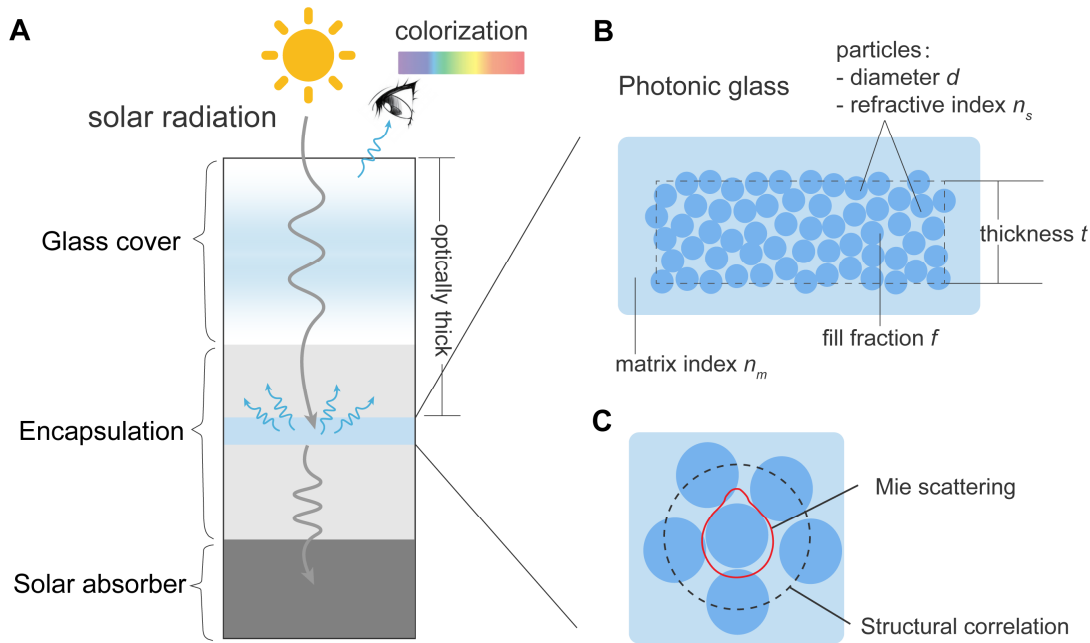

**Figure 1**. (A) Schematic diagram of using a thin film of photonic glass encapsulated in polymers to colorize solar energy harvesting materials. (B) Schematic showing a photonic glass and the parameters that determine its optical properties. (C) The light scattering from a phonic glass assembly could be regarded as a superposition of Mie scattering by a single particle and interference of the light scattered from correlated particles.

As a colored cover for solar energy harvesting materials like solar cells that work outdoor, the photonic glass layer is better to be embedded in a polymer encapsulant (**Figure 1A**). Additionally, there should be a glass cover to resist environmental damage. In this context, the matrix of the photonic glass is a polymer with a refractive index from 1.4 to 1.6, rather than the commonly used air. In the analysis of this study, we take both the glass cover and the polymer encapsulant as an optically thick layer with a refractive index of 1.5. Besides, in order to obtain a certain degree of selective reflection with a small quantity of materials, it is preferable to use particles with a high refractive index. The photonic glass in this case has rarely been specifically discussed before. Furthermore, because of the disordered structure and the strong multiple scattering, accurate modeling of the photonic glass remains a critical challenge.

Here, we use full-wave electromagnetic simulation in 3D dimensions to study the optical reflection and transmission of thin films made by photonic glasses with relatively large refractive index contrast. Through the simulated spectral reflectance in broad wavelengths from UV to NIR, we have evaluated the generated color and average solar transmissivity (AST), as well as how the particle diameter and layer thickness influence them. Besides, we have examined some potential enhancement methods, such as increasing the fill fraction, improving the structure factor, and using core-shell structures or air for scattering particles. Furthermore, we provide explanations about the physical mechanism behind the inability of photonic glasses to achieve highly saturated colors as well as yellow and red hues. Finally, we demonstrate that by adding absorptive materials or forming long-range structural order, the multiple scattering in photonic glasses could be eliminated, and thus red color could be realized.

**Simulation**

As shown in **Figure 1B**, the thin film of photonic glass is self-assembled by monodisperse non-absorbing microspheres with diameters of $d$ and refractive indexes of $n_s$. As mentioned above, the matrix is assumed to be a polymer with a refractive index of $n_m$, and $n_m$ is constant at 1.5 in this study. Besides, the fill fraction ($f$) of microspheres in the photonic glass layer and the layer thickness ($t$) would also affect the spectral reflectance. Totally there are five parameters determining the interaction with light, i.e., $d$, $n_s$, $n_m$, $f$, $t$. In this study we have evaluated how these parameters affect the appeared color and solar transmission.

Existing theories describing the light-matter interaction in photonic glasses are mainly based on the diffusion theory of light-scattering media.[27-29] As illustrated in **Figure 1C**, the theory treats light scattering of a phonic glass assembly as a superposition of two processes: Mie scattering by a single particle, and interference between the light scattered from short-range correlated particles.[30, 31] A major approximation is that the scattering process happens in an effective media with a refractive index of $n_{eff}$. Although this theory could capture the resonance peaks in a photonic glass, as we will show in the following, it neglects the near-field effect and multiple scattering, especially when dealing with a relatively large refractive index contrast. Recently Aubry et al.[32] used energy-density coherent-potential approximation (ECPA) for calculating $n_{eff}$, and Hwang et al.[33, 34] proposed a Monte-Carlo model accounting for multiple scattering. Nevertheless, these modeling methods are not well suited to the photonic glass system studied in this paper.

In this context, for quantitatively modeling a thin film of photonic glass assembled by non-absorbing particles and with a relatively large refractive index contrast, here we adopt full-wave numerical simulation that is first-principle. Since it is a broadband simulation, and the size of the simulation object is about 10 times relative to the light wavelength, we used the finite difference time-domain (FDTD) method with 3D simulation dimensions.[35-38] During the simulation, we first used a force-biased algorithm to generate a random close packing of monodisperse hard spheres, with specific $d$ and $f$.[39] Three-dimensional spatial coordinates of each sphere were then imported into the FDTD simulation region to generate the simulation structure (**Figure 2A**). Periodic boundary conditions were adopted in the *x* and *y* directions, and it should be noted that the imported structure also has periodic boundaries (refer to *Supporting text A* and *Figure S1* for details).

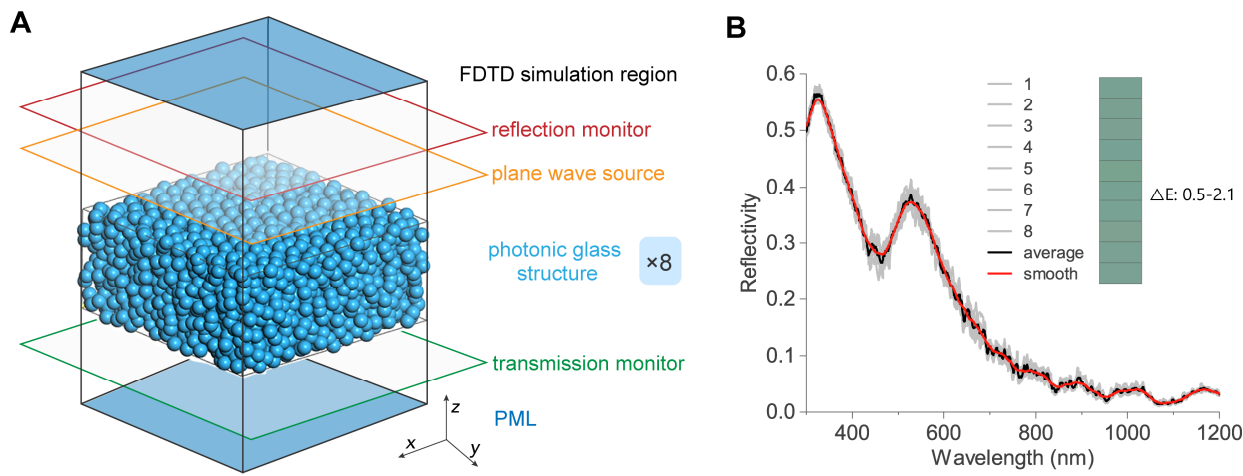

**Figure 2**. (A) Figure illustration of the FDTD simulation region. The boundary conditions at top and bottom *x-y* planes are perfect matched layers (PML), and at four side planes are periodic. A plane wave light source is placed above the photonic glass structure, and the light propagates along the *-z* direction. Frequency domain field profile monitors are placed above the light source and below the structure to measure reflectance and transmittance, respectively. (B) Simulated reflectivity curves of a typical thin film of photonic glass, including results of 8 randomly generated structures (marked as 1-8, grey lines), the average result (black line), and the smoothed curve of the average result (red line). Colors generated by these reflectivity curves are also presented, with a small color difference (ΔE) between each of them.

But since photonic glasses do not have real periodicity, this approach actually simulates a quasi-periodic photonic glass film with the simulation domain as the basic unit. This is somewhat different from actual situations. Therefore, to reduce simulation errors, it is necessary to have enough large simulation regions and to simulate more randomly generated structures. Accordingly, the length of each simulation structure in the *x* and *y* directions was ensured to be larger than 3 μm. Besides, for each simulation case, we simulated 8 randomly generated structures, and make the average as the final result. As shown in **Figure 2**B, the reflectivity curve of each simulated structure shows a similar shape but exhibits some small fluctuations and thus poor smoothness. By taking the average, the reflectance curve becomes much smoother. Furthermore, to better show the main features of the reflectivity curve, we have smoothed it using the Savitzky-Golay method. It should be noted that the colors generated by these reflectivity curves in **Figure 2**B are difficult to be distinguished by naked eyes, with the color difference (ΔE) between the smoothed curve and original curves lower than 2 (see *Supporting text B* for the calculation about color). In the following, the same treatments are adopted for all the simulation cases, and only the final smoothed curves will be presented. Please be noted that the final results will also consider the light reflection at the air-glass interface (see *Equation S2*).

## Results and Discussion

### Comparison between air matrix and polymer matrix

At first, we investigated the effects when changing from an air matrix with $n_m$ of 1.0 to a polymer matrix with $n_m$ of 1.5, by modeling the photonic glasses with $f$ of 55% and $t$ of 3 μm (**Figure 3**). In the air matrix, we use particles with $n_s$ of 1.5, close to that of commonly used colloidal microspheres like $SiO_2$, PMMA, and PS. As shown in **Figure 3A**, by varying particle diameters ($d$), the reflection peaks move over the VIS wavelength range ($\lambda_0$ from 400 to 700 nm), i.e., $\lambda_0$=424 nm @$d$=200 nm, $\lambda_0$=529 nm @$d$=250 nm, and $\lambda_0$=636 nm @$d$=300 nm. This fully demonstrates the structural resonance in the photonic glass. Since the reflection peaks locate at wavelengths corresponding to blue, green, and red spectrums, respectively, the standard sRGB colors generated by these reflectivity curves are greatly varied (displayed by the color rectangles in **Figure 3A**). However, due to the strong reflection at short wavelengths, the color generated by the photonic glass with $d$ of 300 nm appears closer to purple rather than the expected red.

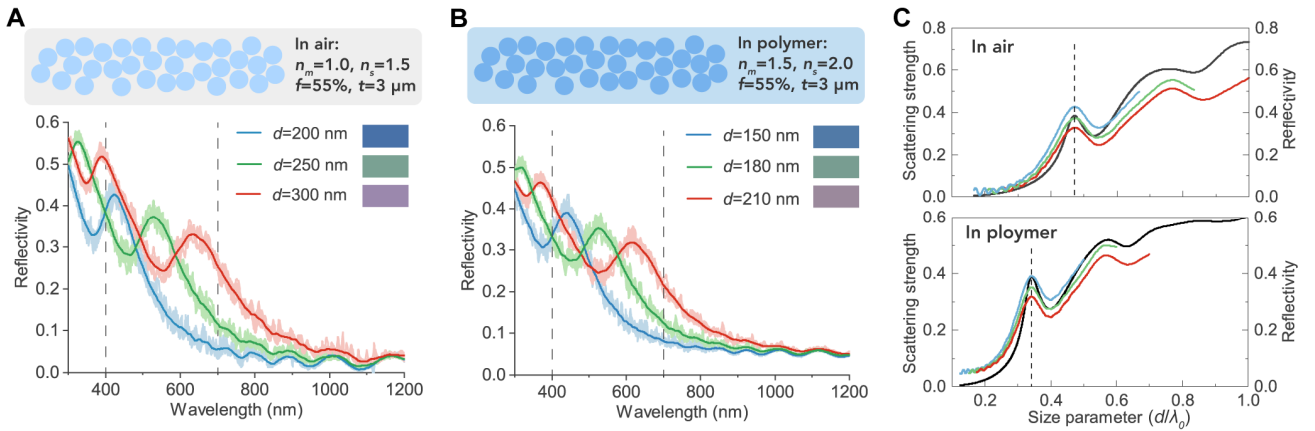

**Figure 3**. Simulated spectral reflectivity for a photonic glass thin film (A) in an air matrix and (B) in a polymer matrix. The air matrix has a refractive index ($n_m$) of 1.0, and the particle refractive index ($n_s$) is 1.5. For polymer matrix $n_m$ is 1.5, and $n_s$ is 2.0. We assume the refractive index has no dispersion and the imaginary refractive index is 0. By varying particle diameters $d$, the reflection peaks are tuned over the VIS wavelength range and thus different colors could be generated, which are displayed by the color rectangles. (C) The reflectivity curves replotted as a function of the size parameter, which is the particle diameter divided by wavelength ($d/\lambda_0$). The scattering strength calculated using the single-scattering model is also plotted by the black lines (refer to *Supporting text C*). All the simulated photonic glasses have a fill fraction of 55% and a thickness of 3 μm.

In order to make the photonic glass in a polymer with scattering intensity similar to that in air, $n_s$ must also be increased accordingly (see *Figure S6* for the effects of $n_s$). Therefore, here we assume that the particles have a $n_s$ of 2.0 when they are in the polymer matrix, and by doing so we obtain very similar results (**Figure 3B**) to those presented in **Figure 3A**. Nevertheless, it should be noted that the particle diameters are notably reduced. The reflection peaks appear at: $\lambda_0$=439 nm @$d$=150 nm, $\lambda_0$=525 nm

@$d$=180 nm, and $\lambda_0$=613 nm @$d$=210 nm. This phenomenon could be explained by the diffusion theory for light-scattering media, which points out that the structure-correlated resonance peak is determined by:[30]

$$\lambda_{resonance} = \frac{4\pi n_{eff} d_{avg}}{x_0} \sin(\theta/2) \qquad (1)$$

where $n_{eff}$ is the effective refractive index of the medium, $d_{avg}$ is the average center-to-center spacing between nearest particles, proportional to $d$. For backward scattering, the scattering angle $\theta$ is $\pi$ and $\sin(\theta/2)$ equals 1. Besides, $x_0$ is the location of the first peak in the structure factor.[30] Therefore, for photonic glasses with the same structure factor, the reflection peak positions are determined by $n_{eff} \cdot d$. With $n_{eff}$ increasing from the air matrix to the polymer matrix, $d$ should thus be reduced to keep $\lambda_{resonance}$ unchanged.

To better explain the simulation results, we have replotted the reflectivity curves as a function of size parameter, which is defined by the particle diameter divided by wavelength ($d/\lambda_0$). As presented in **Figure 3C**, the reflectance curves show almost the same shape no matter $d$, with the same resonance peak positions. For higher $n_{eff}$ in the polymer matrix, the peak positions move to the left. Furthermore, we calculate the scattering strength (refer to *Supporting text C*), presented by the black lines in **Figure 3C**, which exhibit a pretty similar shape and nearly the same response peak positions as the reflectivity curves. Here to promise the peak positions are the same, we use a $n_{eff}$ of 1.24 with respect to the air matrix case, which is a little smaller than those calculated using the classical methods, such as volume average, Maxwell-Garnett approximation, and Bruggeman approximation.[31] So as the polymer matrix case, in which we use a $n_{eff}$ of 1.71. Thus, we validate that with appropriate evaluation of $n_{eff}$, the diffusion theory could capture the main features of light scattering in photonic glasses.

**Effect of thickness on reflectance, color, and solar transmissivity**

Based on the results presented in **Figure 3B**, we extend the thickness of phonic glass films to investigate how it affects spectral reflectance, color appearance, and solar transmittance. We have studied the thickness of 1, 3, 5, and 7 μm, with other parameters unchanged. The results for $d$ = 150 nm, 180 nm, and 210 nm are presented in **Figure 4A**, which clearly show that the reflectance increases with thickness, while the increase rate is gradually slowed. Since increasing thickness would lead to a significant increase in computational effort, we did not simulate the photonic glass films with more thickness values. But based on existing results, it is not difficult to extrapolate the reflectance at higher thicknesses. More critically, it is found that the inverse of transmissivity, i.e., 1/(1-R), at structural resonance peaks shows an almost linear relationship with the thickness (**Figure 4B**), as follows.

$$\frac{1}{1-R} \propto t \qquad (2)$$

This relationship also applies to the reflectivity at other wavelengths (*Supporting text D* and *Figure S2*).

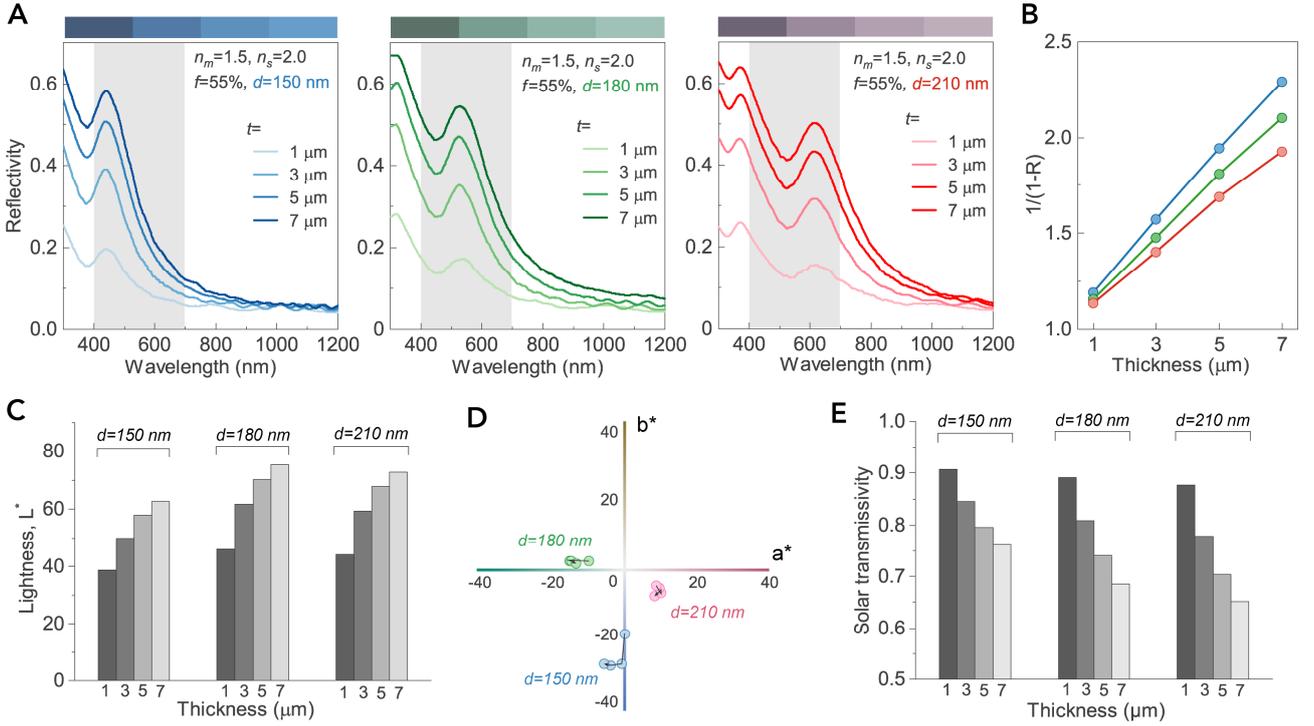

**Figure 4**. (A) Spectral reflectivity (*R*) at different photonic glass film thicknesses (*t*). From left to right, the particle diameters are 150 nm, 180 nm, and 210 nm, respectively. (B) The inverse of transmissivity, i.e., 1/(1-R), at structural resonance peaks as functions of *t*, showing an almost linear relationship. Comparison of the (C) color lightness (CIE $L^*$), (D) color hues (CIE $a^*$ and $b^*$), and (E) the average solar transmissivity.

In **Figure 4A**, the sRGB color for each spectral reflectance curve is presented by the rectangular filled with the corresponding color. With increasing thickness, the appeared colors remain similar hues while being brighter and less saturated. To quantify the colors that photonic glasses can exhibit and clarify their variation with structural parameters like *d* and *t*, we further calculate the CIE $L^*a^*b^*$ color coordinates for each reflectance curve (see the introduction about $L^*a^*b^*$ color space in *Supporting text B*). As shown in **Figure 4C**, the color lightness rises with increasing film thicknesses, keeping a similar pace with reflectivity. However, the color hues are not greatly influenced by the thickness, especially when the film thickness is larger than 3 μm, as demonstrated in **Figure 4D**, further increasing the thickness does not significantly alter the positions on the chromaticity chart. Thus, the relative saturation $C^* = \sqrt{a^{*2} + b^{*2}}$ will only be slightly changed from a thickness of 3 μm to 7 μm. But because color lightness is improved, the appeared colors are more unsaturated. Besides, the results demonstrate that photonic glasses could enable blue colors to be relatively saturated, but the relative saturation for green colors is only half, around 15. As mentioned above, for large particles that make reflection peaks at over 600 nm, the generated color is purplish red instead of pure red, and the saturation is also very low.

To be applied for colorizing PVs and STs, the transmittance of solar energy is a critical judgment criterion. Here we calculate the average solar transmissivity (AST) in wavelengths range of 300-1200 nm (see *Supporting text E*), and the results are presented in **Figure 4E**. Since the reflectivity should be very low in wavelengths beyond 1200 nm, we expect the AST will be higher when considering the whole solar spectrum range. As can be seen, when using small particles to obtain blue color hues, the resultant AST is higher, about 84.6% at lightness of 50. Even for a thickness of 7 μm, over 76% of solar radiation could be transmitted. Increasing particle diameters leads to the reduction of solar transmissivity. For example, at a thickness of 3 μm, the AST is 84.6% for 150 nm, 80.1% for 180 nm, and 77.9% for 210 nm. In brief, we suggest that photonic glasses are suitable for achieving blue hues and can achieve appealing colors, close to sky blue, and also maintain high transmittance. For unsaturated green color hues, using photonic glasses is also a viable choice, which could transmit over 80% of solar radiation when generating a green color with a lightness of ~62. However, it seems that red color hues cannot be achieved with photonic glasses, and using large particles could not bring satisfactory results.

**The effects of potential improvement methods**

In this section, we have studied the effect of several parameters, including structure factor, fill fraction, and scattering particle types, and have checked if there are possible methods to increase color purity and AST. First, based on the random close packing of spheres that are generated by the Force-biased algorithm, we used a Monte-Carlo method and a modified Weeks-Chandler-Anderson (WCA) potential to lower its energy (see details in *Supporting text F*).[34] After the simulation, the improved structure exhibits a higher and narrower peak in its structure factor (calculated by *Eq. S18*), demonstrating an improvement in the short-range order. This results in a slightly more concentrated reflectance curve (**Figure 5A**), with the peak a little bit higher and the valley lower. Thus, the generated color has slightly higher saturation and increased solar transmissivity, but the degree of improvement is extremely limited (**Table 1**). In this case, it is suggested that a photonic glass structure with higher short-range order is preferred in practice, but meanwhile, such an improvement would not lead to a significant change so as to achieve red hue.

Another way to change the structure factor is by varying the fill fraction of particles. Compared to a fill fraction of 55% used previously, here we generate the photonic glass with a fill fraction of 64%, which is the maximum value for a random close packing of hard spheres.[40] As shown in **Figure 5B**, when the structure has a higher fill fraction, the first peak of the structure factor gets higher and narrower. Besides, the peak position is shifted slightly to the right. Therefore, to promise the peak positions of spectral reflectance curves almost unchanged, the particle diameter is increased from 210 nm to 212 nm (**Figure 5B**). In addition, the spectral reflectivity decreases in all the wavelengths. Consequently, to obtain the same reflection peak height, we increase the film thickness from 3 μm to 3.4 μm. We attribute this decrease in reflectance to the weakening of scattering intensity. Finally, the final reflectance curve obtained has a slightly sharper reflectance peak, which is mainly due to its faster drop toward NIR wavelengths (**Figure 5B**). As result, color saturation improved by 0.3, and the average solar transmissivity increases from 77.9% to 79.2% (**Table 1**).

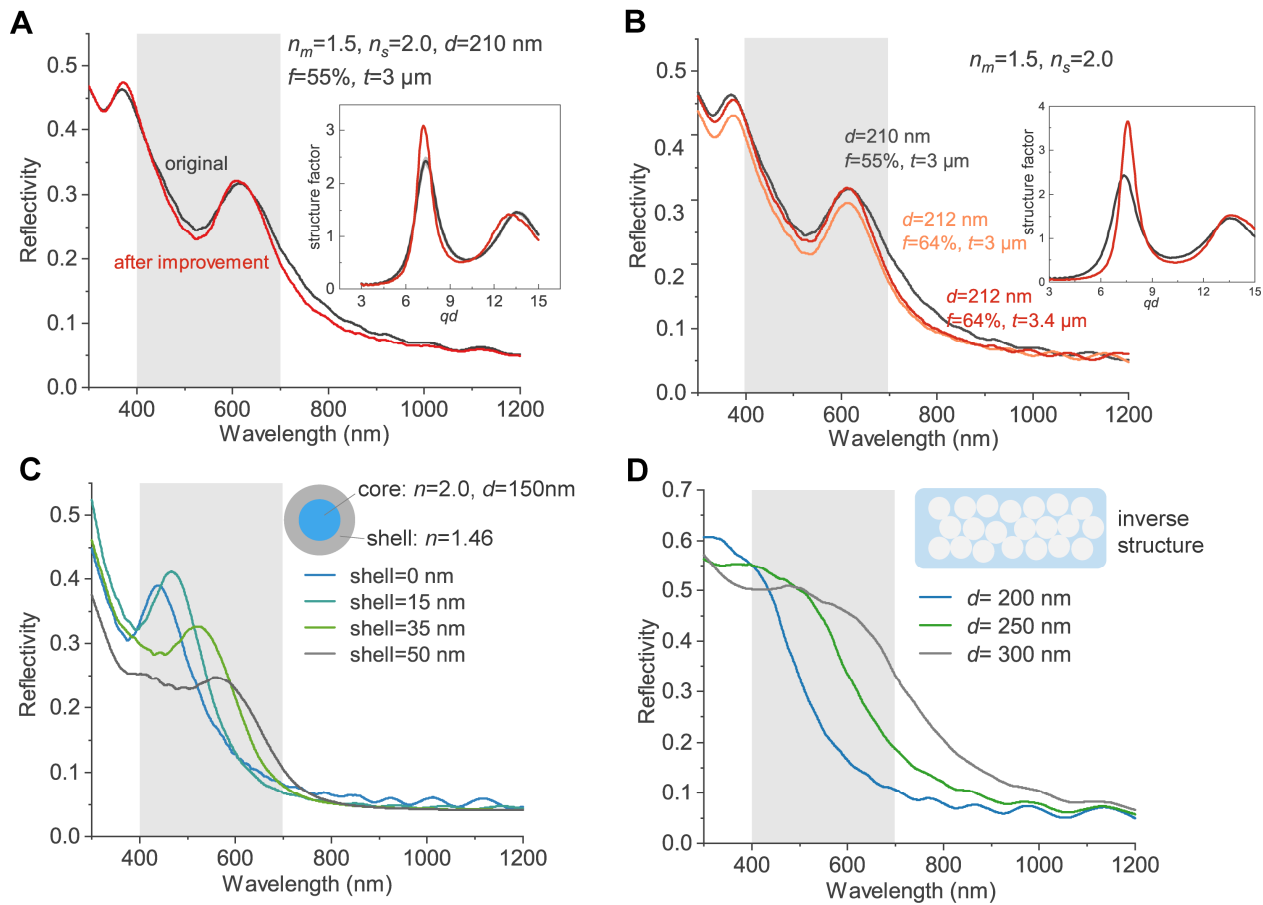

**Figure 5**. (A) The spectral reflectance for the photonic glass with improved structure factor. The insert figure compares the structure factor with/without improvement, as functions of $qd$, with $q=2k\sin(\theta/2)$ and $k=2\pi n_{eff}/\lambda_0$. All the other parameters for these two structures are the same. (B) The effects of increasing fill fraction. With the fill fraction increasing from 55% to 64%, the particle size and film thickness need to be increased to guarantee the same reflection peak position and reflection peak height. (C) Simulated spectral reflectance for photonic glasses made of core-shell particles with different shell thicknesses. The core has a diameter of 150 nm and a refractive index of 2.0. The shell has a refractive index of 1.46. (D) Simulated spectral reflectance for inverse photonic glasses with varying particle diameters, using air as the material of scattering particles.

Table 1. The details about the generated color and AST corresponding to those spectral reflectivity curves presented in Figure 5.

| Methods | Structure parameters | Lightness, L* | Relative saturation, C* | sRGB | AST |
|---|---|---|---|---|---|
| Improving structure factor | $d$=210 nm, $f$=55%, $t$=3 μm | 59.4 | 12.4 | 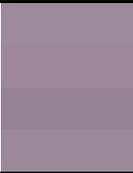 | 0.779 |
| | after improvement | 58.8 | 13.8 | | 0.788 |
| | $d$=212 nm, $f$=64%, $t$=3 μm | 56.8 | 13.2 | | 0.805 |
| | $d$=212 nm, $f$=64%, $t$=3.3 μm | 58.8 | 12.7 | | 0.792 |
| Core-shell particles | $d$=150 nm (shell=0 nm) | 50.0 | 28.6 | 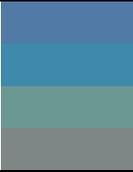 | 0.846 |
| | shell=15 nm | 54.1 | 28.0 | | 0.837 |
| | shell=35 nm | 59.6 | 16.3 | | 0.835 |
| | shell=50 nm | 55.5 | 3.3 | | 0.851 |
| inverse structures | $d$=200 nm | 54.8 | 30.7 | 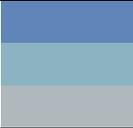 | 0.798 |
| | $d$=250 nm | 70.7 | 15.6 | | 0.716 |
| | $d$=300 nm | 74.5 | 4.2 | | 0.660 |

Using core-shell particles is a prevailing strategy in practice, which can modify particle surface properties and modulate reflective colors. Here, based on the photonic glass made by particles with $d$ of 150 nm and $n_s$ of 2.0, we simulate the cases when the particles are coated with different thicknesses of shells, which have a refractive index of 1.46, close to that of silica. Since the refractive index difference between the shell and the matrix is pretty small, only the core is responsible for scattering light. But higher shell thickness makes $d_{avg}$ increase, which will enable the structural resonance peak to right-shift, as demonstrated by the results in **Figure 5C**. Consequently, diverse color hues including blue, cyan, green, and grey are obtained by using the same core particles (see **Table 1**). Meanwhile, it is found that the reflectivity first gets higher when the shell is thin and then falls with a thicker shell (**Figure 5C**). When the thickness is 15 nm, the fill fraction of the core with $n_s$ of 2.0 is reduced to about 32%. Therefore, it is believed that for a photonic glass structure, there should be an optimal fill fraction that maximizes the light scattering intensity and thus reflectivity. Meanwhile, a higher fill fraction is beneficial for a sharper reflection peak, thus there is a trade-off to be made regarding the fill fraction. When the shell thickness is 50 nm, the structural correlation is so weakened that the corresponding reflection peaks almost disappear (**Figure 5C**). Thus, it is found that using core-shell particles has its unique role but also fails to achieve a red hue.

Finally, we replace the high refractive index particles with the particles using the lowest refractive index material, i.e., air ($n_s$=1.0), resulting in the photonic glass with an inverse structure (**Figure 5D**). Such an inverse structure could be made by using hollow microspheres or etching the encapsulated particles in a polymer.[41, 42] Since this reduces $n_{eff}$, particles with larger diameters should be used. Thus, we simulate the cases with $d$ of 200 nm, 250 nm, and 300 nm, respectively. However, the results as shown in **Figure 5D** are very different from what is presented in **Figure 3A-B**, although the simulated photonic glasses differ only in the material refractive index. Surprisingly, the structural resonance peaks cannot be identified from the spectral reflectivity curves. If further increasing the fill fraction to improve the structure factor peak, we can also observe only small fluctuations in the reflectance curve (*Figure S7*). Therefore, using inverse structures could enable blue colors, and also light cyan colors rather than green. Moreover, we find that using 300 nm particles leads to a nearly neutral color with lightness being 74.5 and relative saturation being 4.2 (**Table 1**). This inspires us that inverse photonic glasses can be used as VIS reflectors with high NIR transmission, enabling a nearly white appearance (see also *Figure S7*).

**Discussion of the intrinsic limitations and solutions**

Until now we have investigated the basic optical properties of photonic glass thin films and some modification methods that may yield some improvement or different properties. However, using the proposed photonic glass technology, we found that yellow and red color hues were still not achievable under the following two strict conditions: 1) the particles are assembled in an amorphous structure with only short-range order; 2) the particles and matrix are made of dielectric materials and are nearly non-absorbing for solar radiation. We suppose that only when either rule is broken, more color hues could be achievable.

Intuitively, it is the high reflectivity at short wavelengths that makes the generated color not pure. As pointed out by Hwang et al.,[43] the increase in scattering intensity toward shorter wavelengths is because of multiple scattering, which originates from the scattering of individual particles. Thus, a higher scattering cross section of a single particle will make the multiple scattering stronger. Unfortunately, when intending to produce structural peaks in wavelengths corresponding to red (600-650 nm), the particle diameter should generally be from 190 nm to 310 nm, considering that $n_{eff}$ is generally in the range of (1.2, 1.8). For these scattering particles or particles with even smaller sizes serving as the scattering core, the scattering strength is always higher at short wavelengths. Therefore, it is not feasible to eliminate this multiple scattering by modifying the scattering microspheres, unless using absorptive materials to absorb the light that is multiple scattered. This has been widely used in previous studies to enhance color saturation.[37, 44, 45]

To prove this idea, here we simulate the photonic glasses that are partially absorptive for visible light. To realize it, we have modified the imaginary refractive index of the matrix material, and we assume it is 1/10 of the measured imaginary refractive index of melanin (see details in *Supporting text G* and *Figure S4*). The real refractive index remains at 1.5. In practice, this could be achieved by homogeneously mixing a small quantity of melanin nanoparticles in the polymer matrix. Compared with the results for the non-absorbing photonic glass presented in **Figure 3B**, the addition of absorbing material causes a significant change in spectral reflectivity. As shown in **Figure 6A**, the strong reflection at short wavelengths is greatly suppressed, making the reflection peak at the long wavelength prominent. Thus, the color produced is no longer purplish and is closer to a dark red hue. What's more, based on the inverse photonic glass presented in **Figure 5D**, with absorptive matrix its appeared color changes from grey to brownish yellow (**Figure 6B**). As demonstrated in **Figure 6C**, the colors produced by non-absorbing photonic glasses are mainly limited in the blue-green-purple region of the chromaticity diagram, and such range could be extended to produce yellow and red when using a light-absorbing matrix. But meanwhile, a significant reduction in light transmission is inevitable because of parasitic absorption (*Figure S5*), making this solution unfavorable when applied for colorizing solar energy harvesting materials.

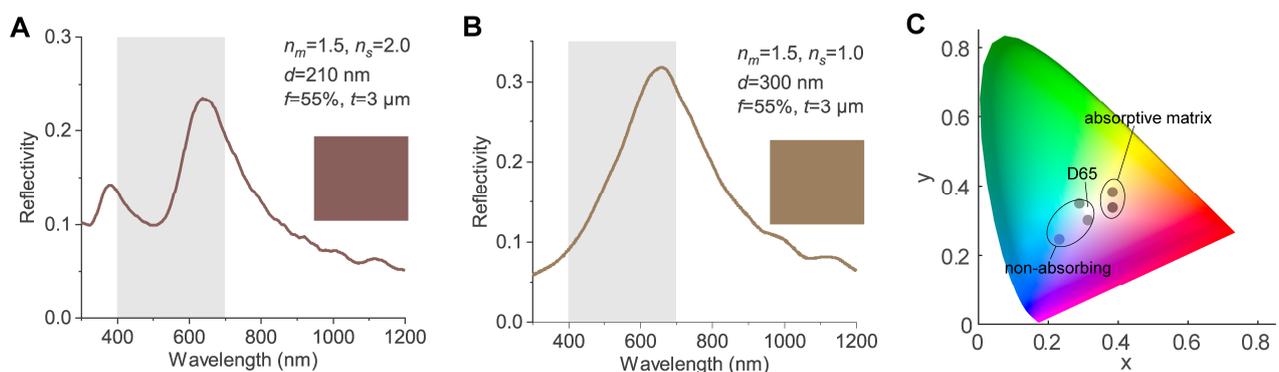

**Figure 6**. Simulated spectral reflectivity of (A) the photonic glass and (B) the inverse photonic glass with a light-absorbing matrix. The colored square displays the sRGB color generated by the reflectance curve. (C) CIE chromaticity diagram. D65: the color of the D65 light source, representing the most neutral color; non-absorbing: the color generated by the non-absorbing photonic glass as shown in Figure 3B; absorptive matrix: the color generated by the photonic glass with the absorptive matrix.

For microspheres that are randomly assembled, it is the impenetrability of hard spheres that imposes a short-range correlation. In reciprocal space, this leads to the structure factor exhibiting a primary peak, as shown in **Figure 7A**. Due to the modification of light scattering by this structural correlation, we can observe the spectral reflection peak at the same location. For the left side of this structural peak, the fast drop of structure factor to nearly 0 leads to a reduction of the scattering efficiency. This is beneficial for the high transparency at long wavelengths (**Figure 7A**) and is desired for solar energy harvesting. However, since particles are randomly packed, for an arbitrary particle we are always able to find other particles at a distance beyond its diameter. As a result, the structure factor oscillates up and down around 1 and would not be close to 0 for the right side where wavelengths get shorter.

Besides, if the first peak of the structure factor locates at red wavelengths, the second peak must locate at UV wavelengths. This means that light scattering in the wavelengths between UV and red spectra cannot be eliminated by the interference effect. Unless the structure factor between the primary and second peak could be down to nearly 0.

Certainly, this requirement could be met by photonic crystals that have long-range order. As demonstrated by **Figure 7B**, the structure factor of an ideal photonic crystal structure is characterized by a series of Dirac peaks in certain positions. This leads to sharp reflection peaks and thus saturated colors. By using particles with $d$ of 200 nm and $n_s$ of 2.0, we show that the photonic crystal structure could generate pretty saturated red hues, even exceeding the domain of sRGB color gamut (**Figure 7B**). Although in practice such a structure with perfect periodicity is unattainable, it is expected that a polycrystalline structure could also provide the realizability of highly saturated colors, which are demonstrated in some recent studies.[25, 46, 47]

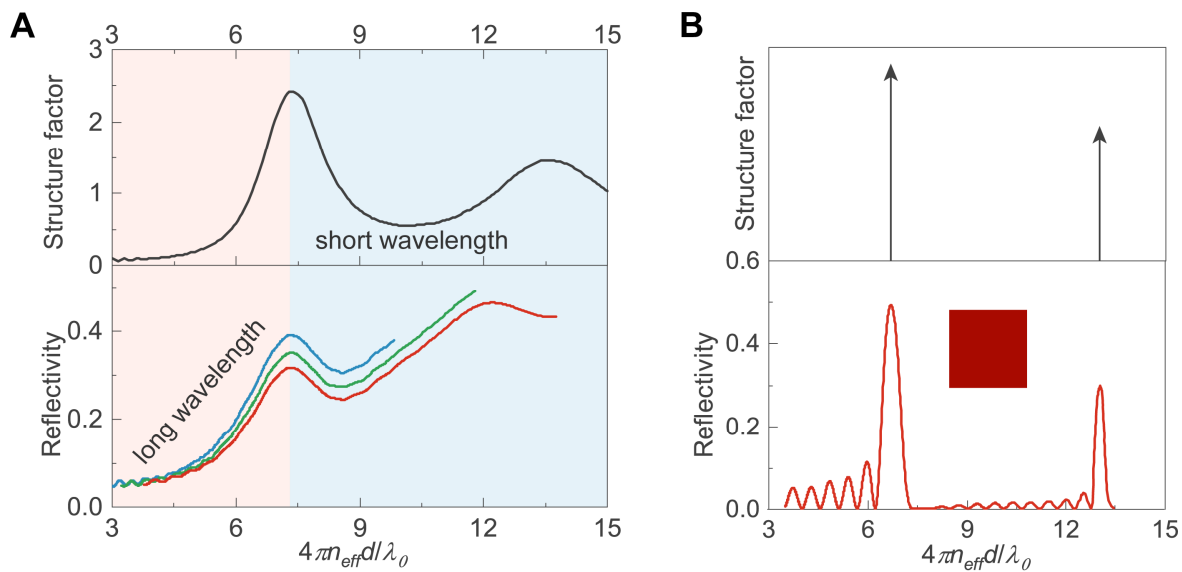

**Figure 7**. The structure factor and spectral reflectivity as a function $4\pi n_{eff}/\lambda_0$: (A) the case of the photonic glass as presented in Figure 3B; (B) the case of a photonic crystal with $d$ of 200 nm and $t$ of 3.1 μm, enabling pretty saturated red color.

**Conclusions**

In summary, this study demonstrates that a thin film made by a random packing of monodisperse dielectric microspheres, i.e., photonic glass, could be a promising candidate for colorizing solar energy harvesting materials. In addition to its advantage of being able to be produced on a large scale and at a low cost, we show that the short-range structural correlation enables selective reflection of VIS light with little NIR reflection. Thus, by using non-absorbing microspheres with relatively high refractive index, we show that a 3 μm thick photonic glass film is capable of producing colors with lightness over 50 yet keeping average solar transmissivity at around 80%. Various color hues can be obtained by changing the microsphere size, while photonic glasses are more recommended to produce hues for only purple, blue, cyan, and light green. Based on both simulation results and theoretical analysis, we validate that the multiple scattering at short wavelengths is unavoidable in a photonic glass structure, resulting in yellow and red color hues unavailable. This limitation could be overcome by either adding light-absorbing materials or using long-range ordered structures to eliminate multiple scattering. Nevertheless, it should be noted by so doing the structure is no longer a photonic glass, and there might be a significant reduction in solar transmittance and an increase in fabrication cost.

We have also studied the effects of some possible improvement methods for photonic glasses. Results suggest that structures with higher short-range order are preferred. This is because the higher and narrower structure factor peak can lead to a narrower reflection peak. Thus, it is essential to optimize the particle assembly conditions in practice. Coating small-size scattering particles with a relatively thin shell may increase the scattering intensity of the photonic glass. Taking advantage of this could be helpful when the core particles are not easily obtainable. When using air as the material of scattering particles, we find that inverse photonic glasses are unexpectedly free of distinguishable reflection peaks, which is greatly different from the results of 2D simulations.[38] Nevertheless, this characteristic makes it feasible for realizing color close to metallic silver or white yet with still over 60% transmissivity of solar energy.

## Materials and Methods

Detailed descriptions of all the methods used in this study are provided in the Supporting Information.

## Author Contributions

Z.L. and T.M. designed the research; Z.L. performed simulations and analyzed data; Z.L. wrote the original manuscript, and S.L. and T.M. revised the manuscript.

## Conflict of Interest

There are no conflicts of interest to declare.

## Acknowledgements

This work is supported by the finical support from the National Natural Science Foundation of China (NSFC) through Grant 51976124 and Shanghai Municipal Science and Technology Commission through Grant 22160713800. The authors thank Anna B. Stephenson for the open-source Python package "structcol" (https://github.com/manoharan-lab/structural-color); and Vasili Baranau for the open-source program "packing generation" (https://github.com/vasilibaranov/packing-generation).

## Supporting information

Supporting information texts: Simulation processes of photonic glass; Calculation about color; Calculation of the scattering strength; Relationship between reflectivity and thickness; Average solar transmissivity; Improving the sphere packing by a Monto-Carlo method; Simulation of photonic glass with light-absorbing matrix. Supporting figures: Figure S1-S6. Supporting references.

**Supplementary Information for**

**Theoretic Guide for Using Photonic Glasses as Colored Covers for Solar Energy Harvesting**

Zhenpeng Li, Sinan Li, Tao Ma*

Engineering Research Centre of Solar Energy and Refrigeration of MOE, School of Mechanical Engineering, Shanghai Jiao Tong University, Shanghai 200240, People's Republic of China

*Corresponding author: Tao Ma (tao.ma@sjtu.edu.cn)

This PDF file includes:

- Supporting information texts

    A. Simulation processes of photonic glass

    B. Calculation about color

    C. Calculation of the scattering strength

    D. Relationship between reflectivity and thickness

    E. Average solar transmissivity

    F. Improving the sphere packing by a Monto-Carlo method

    G. Simulation of photonic glass with light-absorbing matrix

- Supporting figures: Figure S1-S7
- Supporting references

# Supporting information texts

## A. Simulation processes of photonic glass

The simulation process of photonic glass structures is illustrated in Figure S1, including: 1) generation of random hard sphere packings; 2) Dimension conversion of particle coordinates; 3) Establishment of the FDTD simulation file and performing simulation.

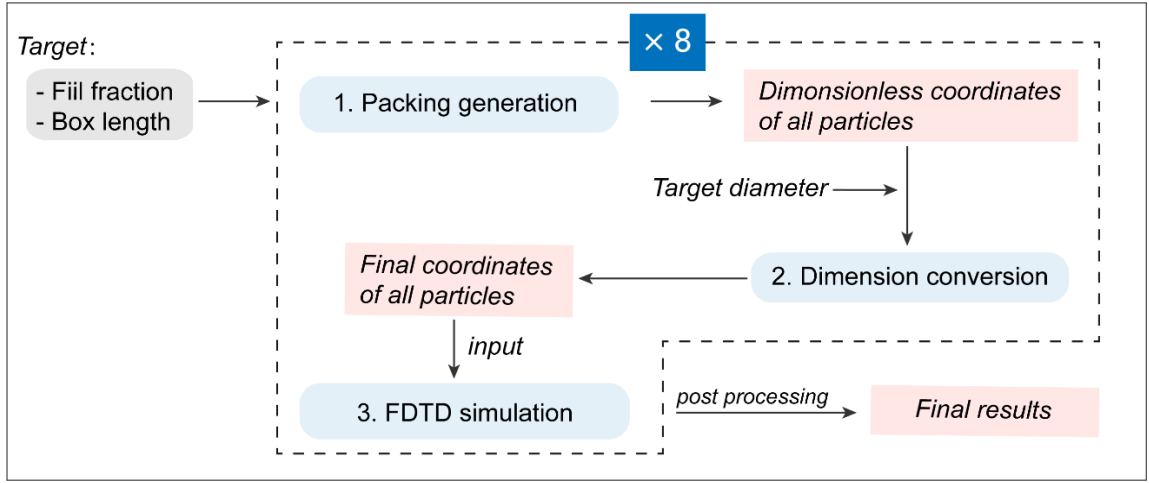

**Figure S1**. FDTD simulation processes of photonic glass.

## 1. Generation of random hard sphere packings

First, we performed numerical generation of random hard sphere packings. We used 'PackingGeneration' program that was developed by Vasili Baranau and distributed on GitHub (https://github.com/VasiliBaranov/packing-generation). Among various generation algorithms, here we used force-biased algorithm that was proposed by Mościński et al.[1] The packing particles reside in a cubic box with periodic boundary conditions in all dimensions. By setting the targe box length ($l$) and fill fraction of particles ($f$), we calculated the number of particles ($N$) based on the following equation.

$$f = \frac{\pi d^3}{6} \frac{N}{l^3} \tag{S1}$$

The suitable value of the Contraction rate was selected to guarantee the final packing has the target fill fraction. Besides, by changing the value of Seed that was used for random number generation, we could get different packings with the same fill fraction and box length yet the different locations of particles.

## 2. Dimension conversion of particle coordinates

After running the program, the 3D spatial coordinates of all particles were read by a MATLAB script. In this script, the coordinates were also converted, making all particles have the target diameter. Generally, after the conversion the box length would be several micrometers. What's more, at the *x* and *y* boundaries of the box, we added a layer of particles outside the box. This promises that when we use periodic boundary conditions in the FDTD simulation region, the simulated structure is strictly periodic. The final spatial coordinates of all particles were saved in a matrix.

**3. Establishment of the FDTD simulation file and performing simulation**

The FDTD simulation was performed by the Lumerical FDTD software. In the simulation file, we first established the photonic glass structure using a script. This script could load the data of particle coordinates and then add each particle at the specified location. All particles were endowed with the target diameter and refractive index. In this study, we used a constant refractive index of 1.5 or 2.0.

For the FDTD simulation region, in the *x* and *y* direction, its length equals the packing box length, and periodic boundaries were applied. In the *z* direction, PML boundaries were applied. The background index was 1.0 (for air) or 1.5 (for polymer). The mesh type was auto non-uniform, and in most cases the mesh accuracy was 3. It was found the simulation results were not very sensitive to the mesh accuracy, which might be attributed to the random photonic structure.

A periodic plane-wave light source was placed above the photonic glass structure, with the light propagating along the -*z* direction. In all the simulation cases, we set the wavelength range as (300 nm, 1200 nm). A 'frequency-domain field and power' monitor was placed above the light source, to record the reflected power and reflectivity. Similarly, another monitor was placed below the photonic glass structure, to record the transmitted power and transmissivity.

**4. Post-processing**

For every simulation case, the above processes were repeated by 8 times. And in each time a different random packing was used for the photonic glass structures. In this case, we would get 8 results for reflectivity, and we would take the average as the final result. Besides, as shown in Figure 2B, to better show the main features of the reflectivity curve, we smoothed the curve using the Savitzky-Golay method.

For the case of air matrix (see Figure 3A), the results of FDTD simulation with a background index of 1.0 were directly used as the final results. For the case of a polymer matrix, based on the results of FDTD simulation with a background index of 1.5 ($R$), we further carried out the following processing:

$$R_{final} = 1 - (1 - R_{air-glass}) \cdot (1 - R) \tag{S2}$$

where $R_{air-glass}$ is the Snell reflectivity at the air-glass interface. In this study we assume both the glass and polymer have a constant refractive index of 1.5. Thus, $R_{air\text{-}glass}$ is 0.04.

## B. Calculation about color

For the calculation of standard colors, the CIE XYZ, xyY, L*a*b* and sRGB color spaces have been used.[2]

**XYZ.** First, the standard color is described in CIE XYZ color space. The X, Y, Z values are calculated by:

$$X = \frac{\int_{\lambda_1}^{\lambda_2} R(\lambda)I(\lambda)\bar{x}(\lambda)d\lambda}{\int_{\lambda_1}^{\lambda_2} I(\lambda)\bar{y}(\lambda)d\lambda} \tag{S3}$$

$$Y = \frac{\int_{\lambda_1}^{\lambda_2} R(\lambda)I(\lambda)\bar{y}(\lambda)d\lambda}{\int_{\lambda_1}^{\lambda_2} I(\lambda)\bar{y}(\lambda)d\lambda} \tag{S4}$$

$$Z = \frac{\int_{\lambda_1}^{\lambda_2} R(\lambda)I(\lambda)\bar{z}(\lambda)d\lambda}{\int_{\lambda_1}^{\lambda_2} I(\lambda)\bar{y}(\lambda)d\lambda} \tag{S5}$$

where $(\lambda_1, \lambda_2)$ is the wavelength range of visible light, $\bar{x}(\lambda)$, $\bar{y}(\lambda)$, and $\bar{z}(\lambda)$ represent the numerical description of the chromatic response of a standard observer (CIE 1931 2°), $I(\lambda)$ is the spectral power distribution of the illuminant, using the standard illuminant D65 in this study.

**xyY.** To be displayed in a 2D chromaticity diagram, CIE XYZ color coordinates need to be converted into xyY values. The $x$ and $y$ are calculated by the following equation.

$$x \text{ or } y = \frac{X \text{ or } Y}{X+Y+Z} \tag{S6}$$

**L*a*b*.** CIE L*a*b* color space is derived from CIE XYZ color space, which is more perceptually uniform. In CIE L*a*b* color space, L* represents perceptual lightness, which defines perfect black at 0 and perfect white at 100. The a* axis is relative to the green-red opponent colors, with negative values toward green and positive values toward red. The b* axis represents the blue-yellow opponents, with negative numbers toward blue and positive toward yellow.

The conversion equations between CIE XYZ and CIE L*a*b* are given as follows.

$$L^* = 116 f\left(\frac{Y}{Y_n}\right) - 16$$

$$a^* = 500\left(f\left(\frac{X}{X_n}\right) - f\left(\frac{Y}{Y_n}\right)\right) \quad \text{(S7)}$$

$$b^* = 200\left(f\left(\frac{Y}{Y_n}\right) - f\left(\frac{Z}{Z_n}\right)\right)$$

with $f(t) = \begin{cases} \sqrt[3]{t} & \text{if } t > \delta^3 \\ \dfrac{t}{3\delta^2} + \dfrac{4}{29} & \text{otherwise} \end{cases}$, $\delta = 6/29$

where $t = \dfrac{X}{X_n}, \dfrac{Y}{Y_n}, \text{ or } \dfrac{Z}{Z_n}$. When using the standard illuminant CIE D65 as the reference white with Y=100, there are $X_n$=95.0489, $Y_n$=100, $Z_n$=108.8840.

**Color difference $\Delta E$.** Generally, the color difference $\Delta E$ is defined in CIE $L^*a^*b^*$ color space, by the following equation.

$$\Delta E = \sqrt{(\Delta L^*)^2 + (\Delta a^*)^2 + (\Delta b^*)^2} \quad \text{(S8)}$$

where $\Delta L^*, \Delta a^*, \Delta b^*$ are the difference in $L^*, a^*, b^*$ coordinates of two different colors.

**Relative saturation $C^*$.** In this study, the relative saturation is evaluated by:

$$C^* = \sqrt{a^{*2} + b^{*2}} \quad \text{(S9)}$$

**sRGB.** To present the color in images, we then convert the XYZ color coordinates to sRGB values, with the following equations.

$$\begin{bmatrix} R_{linear} \\ G_{linear} \\ B_{linear} \end{bmatrix} = \begin{bmatrix} +3.2406 & -1.5372 & -0.4982 \\ -0.9689 & +1.8758 & +0.0415 \\ +0.0557 & -0.2040 & +1.0570 \end{bmatrix} \begin{bmatrix} X \\ Y \\ Z \end{bmatrix} \quad \text{(S10)}$$

$$C_{sRGB} = \begin{cases} 12.92 C_{linear}, & C_{linear} \leq 0.0031308 \\ 1.055 C_{linear}^{1/2.4} - 0.055 & C_{linear} > 0.0031308 \end{cases} \quad \text{(S11)}$$

where C is R, G or B. Finally, the values are multiplied by 255 and round to an integer.

## C. Calculation of the scattering strength

For a light-scattering media, transport mean free path $l^*$ is a measure of the step length for a random walk of photons. Thus, the lower the $l^*$, the higher the scattering strength of light. For the photonic glass that is assembled by monodisperse dielectric spheres, $l^*$ is determined by the following equation.[3, 4]

$$l^* = \frac{1}{1-g} \cdot \frac{1}{\rho \sigma_{sca}} = \frac{1}{1-g} \cdot \frac{\pi d^3}{6 f \sigma_{sca}} \tag{S12}$$

where $\rho$ is the number density of scatters, $f$ is the fill fraction of scatters, $d$ is the scatter diameter. Besides, $g = \langle \cos\theta \rangle$ is the asymmetry parameter and $\theta$ is the scattering angle. $\sigma_{sca}$ is the scattering cross section. They are expressed as:

$$\sigma_{sca} = \frac{\pi}{k^2} \int_\Omega F(\theta) S(\theta) d\Omega = \frac{\pi}{k^2} \int_0^\pi F(\theta) S(\theta) \sin\theta d\theta \tag{S13}$$

$$g = \frac{\int_0^\pi \cos\theta F(\theta) S(\theta) \sin\theta d\theta}{\int_0^\pi F(\theta) S(\theta) \sin\theta d\theta} \tag{S14}$$

where $F(\theta)$ is the form factor, describing the scattering properties of a single scatter (Mie theory);[5] $S(\theta)$ is the structure factor, being the Fourier transform of the pair correlation function of the scattering packings. Generally, it is calculated by using the hard-sphere Percus-Yevick structure factor $S(q)$, with $q=2k\sin(\theta/2)$. $k=2\pi n_{eff}/\lambda_0$ is the wavevector. Therefore, in this theory, all the scatters are assumed to be embedded in an effective medium with a refractive index of $n_{eff}$. As both $F(\theta)$ and $S(\theta)$ depend on $n_{eff}$, the accurate estimation of its value is very important.

In this study, we use $\lambda_0/l^*$ to evaluate the scattering strength of the photonic glass. It should be noted the results presented in Figure 3C are scaled up by 10 times.

## D. Relationship between reflectivity and thickness

After comparing the spectral reflectivity of the photonic glass with different thicknesses, we found that the inverse of transmissivity, i.e., 1/(1-R), has an almost linear relationship with the thickness $t$. In Figure S2, we verified this hypothesis at wavelength 400, 500, and 600 nm. Thus, by only simulating two or three different thicknesses of photonic glass, we may deduce the reflectivity at other thicknesses.

It should be noted that here the $R$ is the direct result of FDTD simulation, without considering the reflectance at the air-glass interface (see Eq. S2).

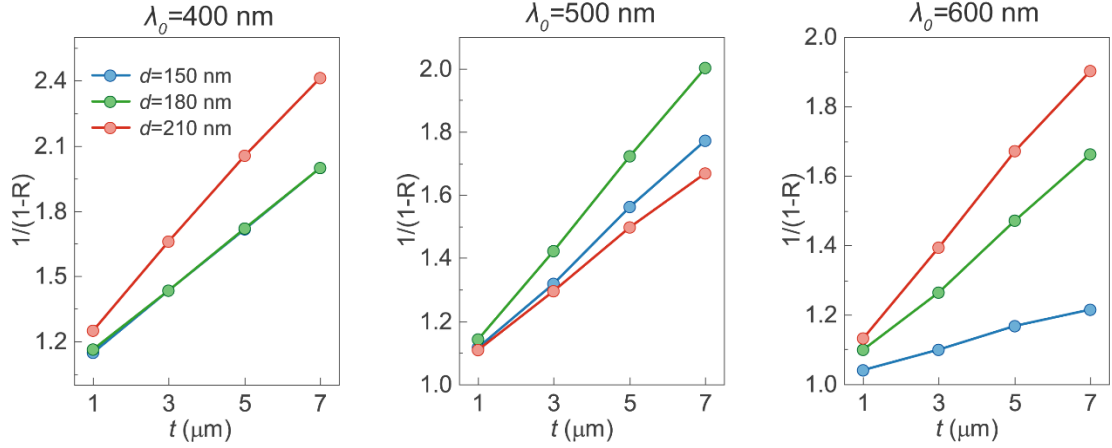

**Figure S2**. The inverse of transmissivity, i.e., 1/(1-R), as functions of thickness *t*. From left to right, the wavelength is 400 nm, 500 nm, and 600 nm. Related to Figure 4B.

### E. Average solar transmissivity

In this study, we use the average solar transmissivity (AST) to evaluate the feasibility of using a photonic glass thin film for colorizing solar energy harvesting materials, calculated by the following equation.

$$AST = \frac{\int_{\lambda_1}^{\lambda_2}(1-R(\lambda))P(\lambda)d\lambda}{\int_{\lambda_1}^{\lambda_2}P(\lambda)d\lambda} \tag{S15}$$

where $\lambda$ is wavelength, $P(\lambda)$ is the spectral irradiance of AM1.5 solar spectrum (see Figure S2), and $R(\lambda)$ is the spectral reflectivity. In this study, we used a wavelength range of (300 nm, 1200 nm), in which the total solar power is about 836 W/m$^2$, accounting for the majority. If designed specifically for solar cells, it is better to use solar photon flux rather than irradiance.

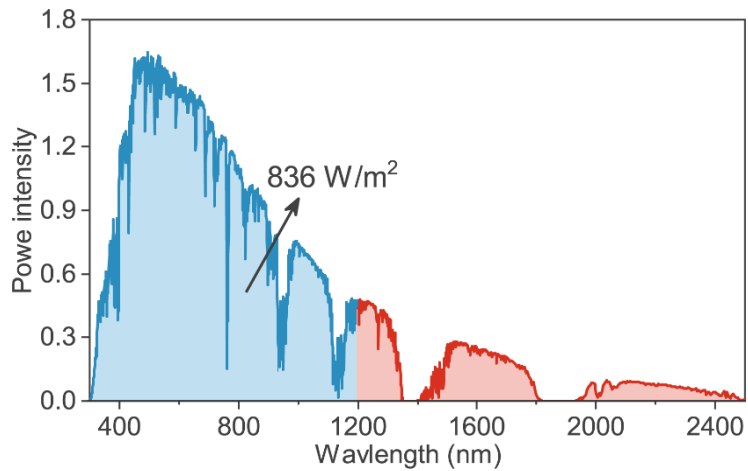

**Figure S3**. AM1.5 solar radiation spectrum. In wavelengths from 300 nm to 1200 nm, the total

power is about 836 W/m².

### F. Improving the sphere packing by a Monto-Carlo method

Based on the sphere packings generated by the force-biased algorithm, we used a Monto-Carlo method to simulate the interactions between spheres and further improve the short-range order. This was carried out by a homemade program in MATLAB.

For two hard spheres, their potential should be described as follows.

$$u(r) = \begin{cases} \infty & r < d \\ 0 & r \geq d \end{cases} \tag{S16}$$

where $d$ is the diameter of the sphere, and $r$ is the distance of the centers of the particles. However, such discontinuous interactions cannot be simulated by molecular dynamic-based forces. In this case, following the work by Jover et al.[6] and Xiao et al.,[7] here we used a modified pairwise Weeks-Chandler-Anderson (WCA) potential, as follows.

$$u(r) = \begin{cases} 4\varepsilon \left[ \left(\dfrac{\sigma}{r}\right)^{200} - \left(\dfrac{\sigma}{r}\right)^{100} + 0.25 \right] & \text{if } r \leq 2^{1/100}\sigma \\ 0 & \text{if } r > 2^{1/100}\sigma \end{cases} \tag{S17}$$

where σ and ε represent the size of the particle and the depth of the potential well, respectively.

In the simulation, the reduced temperature $T^*=k_BT/\varepsilon$ is 1 ($k_B$ is Boltzmann's constant). The cutoff length is 1.1σ. The max displacement of a random move of particles is 0.007σ. Besides, a neighbor-list is used to improve the simulation speed, which is updated after every Monto-Carlo cycle. The simulation starts from the random hard-sphere packing generated by the force-biased algorithm. Each Monto-Carlo cycle contains $N$ random displacements of individual particles, and in this study, we performed 30000 cycles for each structure.

Before and after the Monto-Carlo simulation, the structure factor $S(q)$ was calculated by

$$S(q) = 1 + \dfrac{1}{N}\sum_{j \neq k} \dfrac{\sin(qr_{jk})}{qr_{jk}} \tag{S18}$$

where $q$ is the magnitude of the wavevector, $r_{jk}$ is the radial distance between two particles $j$ and $k$.

### G. Simulation of photonic glass with light-absorbing matrix

To study the effects of adding light-absorbing materials in the photonic glass, we have modified the imaginary refractive index of the matrix material to be nonzero. This was realized by adding a bulk material, which had the same size as the photonic glass structure. Besides, its position coincided perfectly with the photonic glass structure. In this case, the photonic glass was fully surrounded by it. The real refractive index of the bulk material is 1.5, the same as the background index. For the imaginary refractive index, we assume it is 1/10 of the measured imaginary refractive index of melanin. The measured data was taken from Li et al.[8] Figure S3 plots the fitted curve of the imaginary refractive index used for FDTD simulation. Other than that, all other settings remain the same.

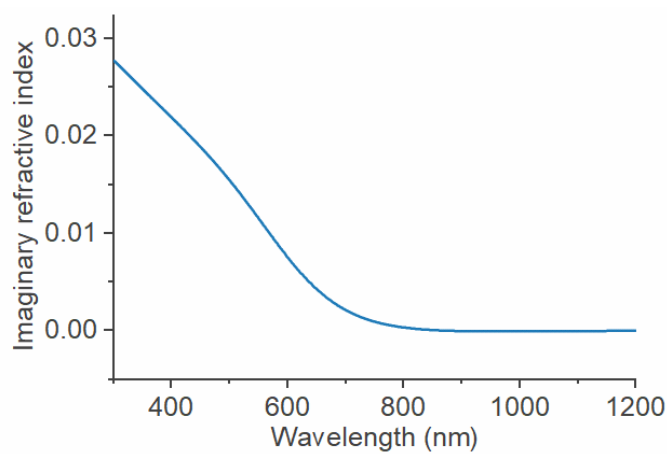

**Figure S4**. The fitted curve of the imaginary refractive index of the matrix material.

Detailed simulation results are presented in Figure S5. As can be seen, the average solar absorptivity is nearly double the average solar reflectivity. In this case, the AST decreases to 0.591 for the photonic glass structure, and 0.474 for the inverse photonic glass structure.

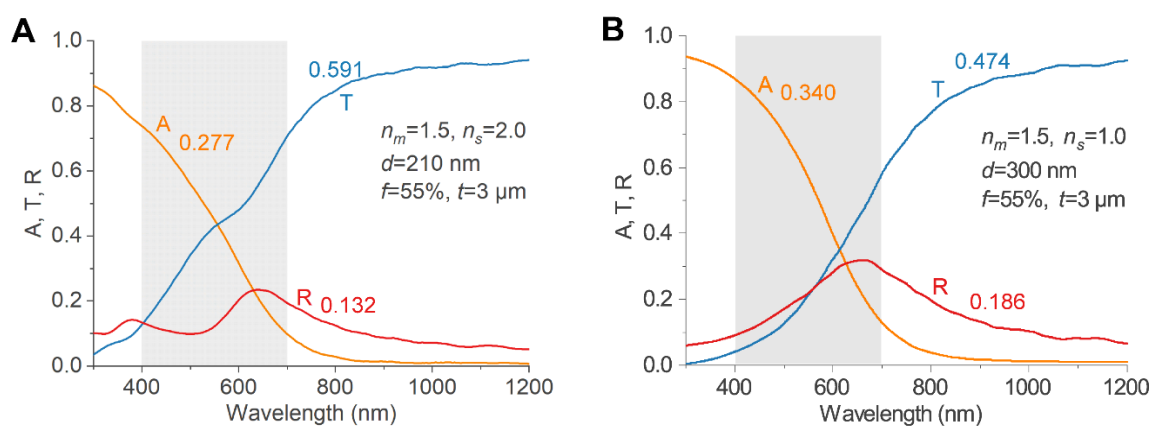

**Figure S5**. Simulated spectral reflectivity $R$, absorptivity $A$, and transmissivity $T$ of (A) the photonic glass structure and (B) the inverse photonic glass structure with absorptive matrix materials. Related to Figure 6.

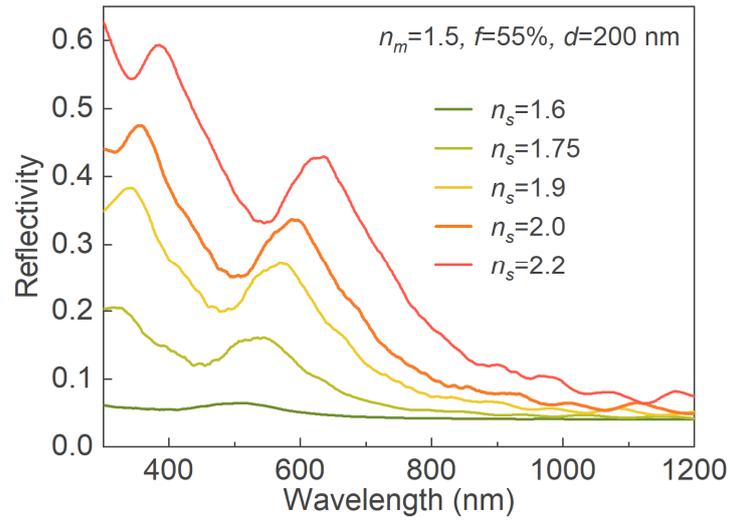

**Figure S6**. The effects of particle refractive index ($n_s$) on the spectral reflectivity of photonic glasses. The matrix refractive index ($n_m$) is 1.5, fill fraction is 55%, particle diameter ($d$) is 200 nm, and thickness is 3 μm.

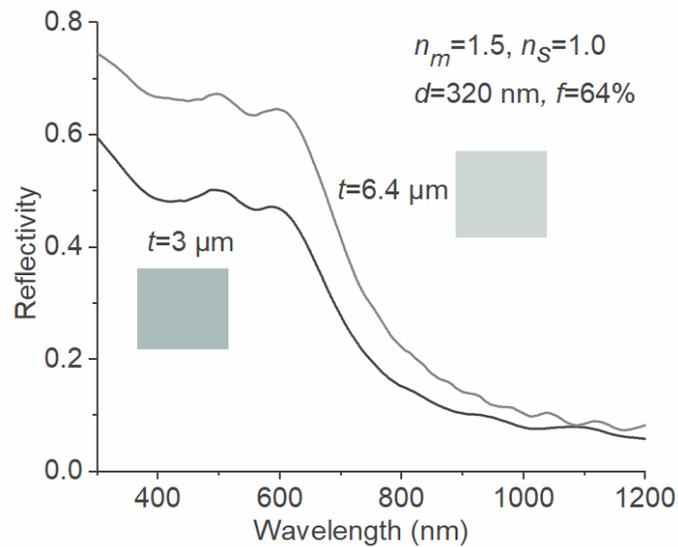

**Figure S7**. Simulated spectral reflectance for inverse photonic glasses, with $f$ of 64% and $d$ of 320 nm. The thicknesses ($t$) are 3 μm and 6.4 μm, respectively.

**Supporting References**